\documentstyle[aps,epsfig]{revtex}
\begin{document}
\draft
\title{Analysis of a continuous field theory in two dimensions 
with use of the Density Matrix Renormalization Group} 
\author{William Lay \cite{newaddress} and Joseph Rudnick}
\address{Department of Physics, UCLA, Box 951547, Los Angeles, CA 
90095-1547}
\date{\today}
\maketitle
\begin{abstract}
A formulation of the Ginzburg-Landau-Wilson version of the partition 
function of a system with a continuously varying order parameter as a 
transfer matrix calculation allows for the application of methods 
based on the Density Matrix Renormalization Group (DMRG) to the 
calculation of the free energy of the $O(1)$ model. The essence of 
both the mapping and the DMRG calculation is laid out, along with 
results that validate this strategy.  This method forms the basis of 
a unified approach to the crossover from three to two dimensions in an 
$O(1)$ system with a slab-like geometry.
\end{abstract}
\pacs{75.40.-s,05.01.-a,64.60.Ak}

The formulation utilized to analyze the critical point of the $O(1)$ 
system in three and more dimensions differs in fundamental respects 
from the formulation that has proven most fruitful in two and fewer 
dimensions.  In the former case, highly accurate results \cite{nickel,LG&ZJ} have been 
obtained with the use of the Ginzburg-Landau-Wilson effective 
Hamiltonian, which is based on a  continuously 
varying order parameter.  On the other hand, lower dimensional systems 
are generally modeled with the use of a fixed-length order parameter 
\cite{lenz,ising}.  

The above state of affairs is acceptable, if not entirely 
satisfactory, as long as one is not interested in a system whose 
critical point properties possesses both three- and two-dimensional 
features.  Such is the case in the instance of dimensional crossover 
in a system, with a slab-like geometry.  Scaling relations and the 
Renormalization Group predict the asymptotic behavior of such a system 
when the correlation bulk three-dimensional correlation length is 
asymptotically large or small compared to the finite slab width.  In 
addition, finite size scaling indicates the existence of a general 
form for the crossover between those two limits \cite{f&b1,f&b2}.  
However, the calculation of the specific form of the crossover 
function requires a unified theoretical approach to the critical point 
properties of a slab-like system in both the three- and 
two-dimensional regimes.

The calculation of the crossover function, and of the thermodynamic 
quantities that can be derived from it, has been carried out in a few 
cases.  For example, crossover from three to two dimensions has been 
worked out in the mean spherical model \cite{f&b1,s&p1} and in its 
mathematical equivalent, Bose-Einstein condensation.  Crossover from 
$d$ to $d-1$ dimensions can also be determined in an $O(1)$, or 
Ising-like, system when $d \ge 5$ \cite{s&p2}.  Here, both systems 
possess asymptotically mean-field critical behavior.

In addition to the above exact determinations of dimensional 
crossover, a field-theoretically motivated approach has been 
formulated based on the notion of an ``environmentally friendly'' 
renormalization group \cite{o&s,os&b}.  This approach yields explicit 
(and relatively simple) crossover functions for thermodynamic 
quantities.  While the functions are necessarily approximate, they are 
consistent with the key expectations that arise from finite-size 
scaling.  Unfortunately, existing literature provides no simple way to 
improve the crossover functions that arise from the environmentally 
friendly renormalization group, so as to produce predictions that can 
be profitably compared with experimental data.

We have developed an approach that holds the promise of producing just 
such an improvement.  This approach is based on a reformulation of the 
partition function of a lattice based system with a continuous order 
parameter in terms of transfer matrices which allows for the 
calculation of the partition function either with the use of 
field-theoretic techniques or with the use of calculational devices 
that have been formulated to speed the numerical evaluation of quantum 
mechanical density matrices in one dimension, a problem that is 
mathematically equivalent to the calculation of transfer matrices in 
two dimensions.

We start with the Ginzburg-Landau-Wilson effective Hamiltonian of an 
$O(1)$ system on a lattice in three dimensions. If $s_{i}$ is the 
value of the continuous spin variable at the $i^{\rm th}$ lattice 
site, then this effective Hamiltonian has the form
\begin{equation}
\frac{1}{2}\sum^{\prime}_{i,j} \left(s_{i}-s_{j}\right)^{2} + 
\sum_{i} \left[\frac{r}{2} s_{i}^{2} + \frac{u}{4} s_{i}^{4} \right]
	\label{effham1}
\end{equation}
The first sum in Eq.\ (\ref{effham1}) is over nearest neighbor pairs.  
High order perturbation theory coupled with the Renormalization Group 
has been shown to yield very accurate results for the 
three-dimensional critical behavior of the system described by the 
effective Hamiltonian (\ref{effham1}).

In two dimensions, the Ising model, which is based on fixed length 
spins, has been successfully investigated with the use of a variety of 
techniques.  An important subset of the approaches is based on an 
analysis of the transfer matrix coupling the state in 
$2^{N}$-dimensional space of a column of spins to the state of a 
neighboring spin column \cite{baxter}.  Onsager's exact solution 
\cite{onsager} for the Ising model partition function consisted of the 
extraction of the largest eigenvalue of this transfer matrix.  In the 
absence of techniques leading to exact solutions, one has recourse to 
other methods for the determination of the largest eigenvalue of a 
two-dimensional transfer matrix.  In particular, there is a set of 
approaches developed to perform highly accurate numerical evaluation 
of the eigenvalues of the density matrix of a one-dimensional quantum 
mechanical system.  The generic term for this collection of 
calculational devices is the Density Matrix Renormalization Group 
(DMRG) \cite{White1}, \cite{White2}. There are cases in which the 
extremely high accuracy of this method has led to important results. 

In the context of statistical mechanics, the DMRG has an analog in the 
form of the Transfer Matrix Renormalization Group (TMRG) \cite{Nishino}. 
The TMRG has been shown to lead to excellent results for the thermodynamic 
functions of the two dimensional Ising model \cite{Nishino},
the $Q$-state Potts model \cite{Carlon1}, \cite{Carlon2} and
confinement effects in the presence of gravity \cite{Carlon3},
\cite{Carlon4}.

We have been able to demonstrate that this method can be applied to 
the evaluation of the partition function for the continuous-spin effective 
Hamiltonian (\ref{effham1}). We start with the 
one-dimensional version of the Ginzburg-Landau-Wilson partition 
function
\begin{equation}
\int \prod_{i}ds_{i} \exp\left[-\sum_{i}\left(\frac{1}{2} 
\left(s_{i}-s_{i+1}\right)^{2} + \frac{r}{2} s_{i}^{2} + \frac{u}{4} 
s_{i}^{4}\right) \right]
	\label{part1}
\end{equation}
This partition function can also be written as the inner product of a 
sequence of transfer matrices $T(s_{i}, s_{i+1})$, where
\begin{equation}
T(s_{i}, s_{j}) = \exp \left[ -\frac{1}{2} \left(s_{i} - s_{j} 
\right)^{2} - \frac{r}{4} \left(s_{i}^{2} + s_{j}^{2} \right) - 
\frac{u}{8} \left( s_{i}^{4} + s_{j}^{4} \right) \right]
	\label{transfer1}
\end{equation}
Eigenfunctions, $\psi_{k}(s)$, of this transfer matrix are defined by 
the equation
\begin{equation}
\int ds^{\prime} T\left(s, s^{\prime} \right) \psi_{k}\left(s^{\prime} 
\right) = \lambda_{k} \psi_{k}(s)
	\label{eig1}
\end{equation}
We will list the eigenvalues in order of magnitude.  The largest 
eigenvalue will be $\lambda_{0}$, the next largest $\lambda_{1}$ and 
so on.  Given that the transfer matrix is real and symmetric, the 
eigenfunctions can also be written as real, normalized and orthogonal, 
in that
\begin{equation}
\int_{-\infty}^{\infty} \psi_{k}(s) \psi_{l}(s) ds = \delta_{k,l}
	\label{orthonorm}
\end{equation}
The one-dimensional transfer matrix is then written as
\begin{equation}
T\left(s, s^{\prime}\right) = \sum_{k} \psi_{k}(s) \lambda_{k} 
\psi_{k}(s^{\prime})
	\label{transfer2}
\end{equation}
Given the orthnormality of the eigenfunctions, the one-dimensional 
partition function has the form
\begin{equation}
T^{N}\left(s,s^{\prime}\right) = \sum_{k} \psi_{k}(s) 
\lambda_{k}^{N} \psi_{k}(s^{\prime})
	\label{part2}
\end{equation}
and is, thus dominated by the contribution associated with the 
largest eigenvalue, $\lambda_{0}$,  of the transfer matrix.

Now, the $d$-dimensional Ising model can be written in terms of the 
eigenfunctions and eigenvalues of the one-dimensional Ising model by 
associating neighboring pairs of spin variables in terms of a ``bond'' 
transfer matrix, $T_{b}(s_{i},s_{j})$, having the form
\begin{eqnarray}
T_{b}\left(s_{i}, s_{j} \right) &=& \exp \left[-\frac{1}{2} 
\left(s_{i}-s_{j} \right)^{2} - \frac{r}{4 d} \left( s_{i}^{2} + 
s_{j}^{2} \right) - \frac{u}{8 d} \left( s_{i}^{4} + s_{j}^{4} 
\right) \right] \nonumber \\
&=& \sum_{k} \psi_{k}\left(s_{i}\right) \lambda_{k} \psi_{k} 
\left(s_{j} \right)
 \label{Tb}
\end{eqnarray}
The partition function is then a sum over bond indices, $k$, while 
vertices contain integrations over the spin variable $s$ of 
combinations of transfer matrix eigenfunctions:
\begin{equation}
\int_{-\infty}^{\infty}\left(\prod_{l=1}^{2d}\psi_{k_{l}}(s) \right) 
ds
	\label{siteint}
\end{equation}
While the eigenfunctions, $\psi_{k}(s)$ are orthonormal, there is no 
reason to believe that this will lead to significant restrictions 
over the allowed indices in the integration of a product of $2d$ of 
the $\psi_{k}$'s when $d>2$. One important simplification arises 
from the fact that a $\psi_{k}(s)$ will be either even or odd in the 
spin variable $s$. In fact, the largest eigenvalue is associated with 
an even-parity eigenfunction and the next largest with an odd-parity 
eigenfunction.

The partition function of the two-dimensional continuous-spin variable 
version of the $O(1)$ model was calculated with the use of the 
formulation above in terms of transfer matrix eigenfunctions, the 
basis set being truncated at the eigenfunctions having the largest few 
eigenvalues.  The fourth order coupling constant, $u$, in 
(\ref{effham1}) was chosen to be equal to 0.23, guided by a best fit 
between the Ginzburg-Landau Wilon effective Hamiltonian and order 
parameter distributions in three dimensions as obtained by simulation 
\cite{RL&J}.  These eigenfunctions were used to construct the basis 
set of the two-dimensional transfer matrix connecting one row of 
continuous spins to the next.  This two-dimensional transfer matrix 
was then reduced with the use of an adaptation of the TMRG. The free 
energy is extracted from the largest eigenvalue of the transfer matrix 
via the connection ${\cal F} = - k_{B}T \ln \lambda_{0}$.  Because we 
are interested in the behavior of the partition function in the 
immediate vicinity of the critical point, temperature-dependent 
factors multiplying the log of the largest eigenvalue are set equal to 
a constant.

This version of the free energy leads immediately to the specific heat 
through differentiation with respect to the ``bare'' reduced 
temperature, $r$.  It is also possible to introduce a 
symmetry-breaking field, $h$, appearing in the effective Hamiltonian 
through the term $-\sum_{i}hs_{i}$. The remanent magnetization and the 
isothermal susceptibility can be calculated from the free energy by 
taking, respectively, first and second derivatives with respect to 
the symmetry-breaking field.

\begin{figure}
    \centering
    \mbox{\epsfig{file=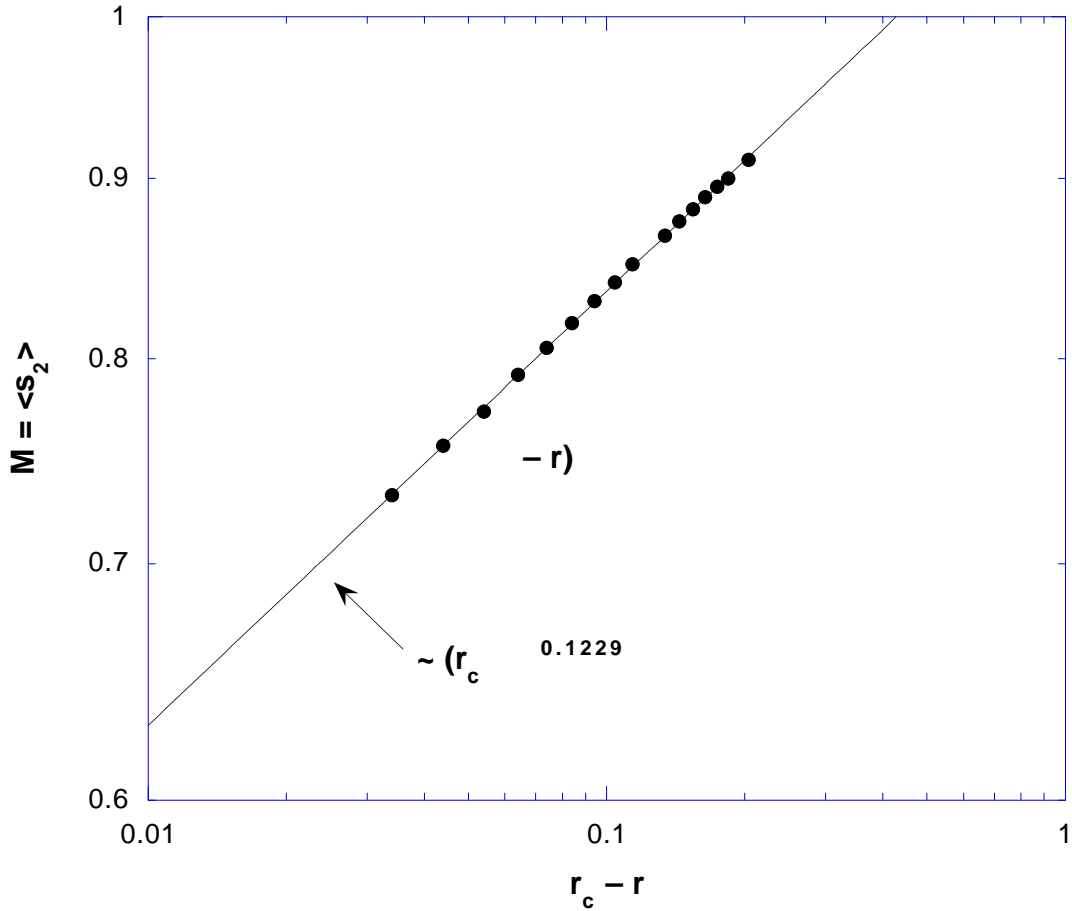,height=4.8in}}
    \caption{The spontaneous magnetization (calculated as the 
    expectation value in the state $|\Psi_{0}\rangle$) 
    as a function of $(r_{c}-r)$.}
    \label{fig:betan2}
\end{figure}

The spontaneous magnetization of the system can be calculated from the 
spin operator $s_{2}$
\begin{equation}
	M=\langle\Psi_{0}|s_{2}|\Psi_{0}\rangle
	\label{spinexpectation}
\end{equation}
where $\Psi_{0}$ is the eigenfunction of the transfer matrix with the 
largest eigenvalue and the spin operator $s_{2}$ is the Ising spin 
operator.  We plot the value of (\ref{spinexpectation}) on a log-log 
grid to illuminate the power-law behavior (see Figure 
\ref{fig:betan2}).  The value of the critical exponent $\beta$ is 
calculated to be $0.1229 \pm 0.0006$ which is consistent with the 
exact value for the $d=2$ Ising model of $1/8$ \cite{baxter}.

\begin{figure}
    \centering
    \mbox{\epsfig{file=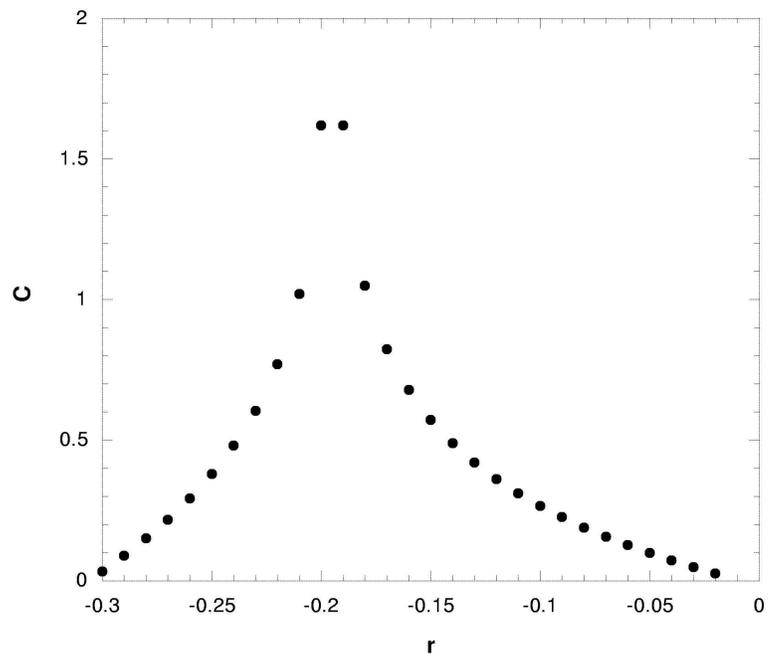,height=4.8in}}
    \caption{The specific heat as a function of $r$.}
    \label{fig:Cn2}
\end{figure}

Differentiating the free energy twice with respect to the reduced 
temperature $r$, we obtain the specific heat.  We display our results 
for this quantity as a function of $r$ in Figure \ref{fig:Cn2}.  
Further analysis of these data reveal a logarithmic dependency both 
above and below the critical temperature, in other words $\alpha = 
\alpha\prime = 0$, as expected from the Ising model.

We have also examined the ratio of specific heat amplitudes 
$A_{+}/A_{-}$ where we fit to the function
\begin{equation}
	C_{\pm}=A_{\pm}\ln |r-r_{c}|+B_{\pm}
	\label{CfitEq}
\end{equation}
For an $O(1)$ system in two dimensions, this value is expected to be 
unity. The value calculated from the data is 0.94.

As mentioned in the previous section, by adding a symmetry-breaking 
term $-\sum_{i}hs_{i}$, we are able to take derivatives with respect 
to $h$ to obtain the spontaneous magnetization (independently from 
the $\langle\Psi_{0}|s_{2}|\Psi_{0}\rangle$ method) and the magnetic 
susceptibility $\chi_{T}$. Determination of the magnetization using 
this approach gives $\beta = 0.1230 \pm 0.0009$. 

\begin{figure}
    \centering
    \mbox{\epsfig{file=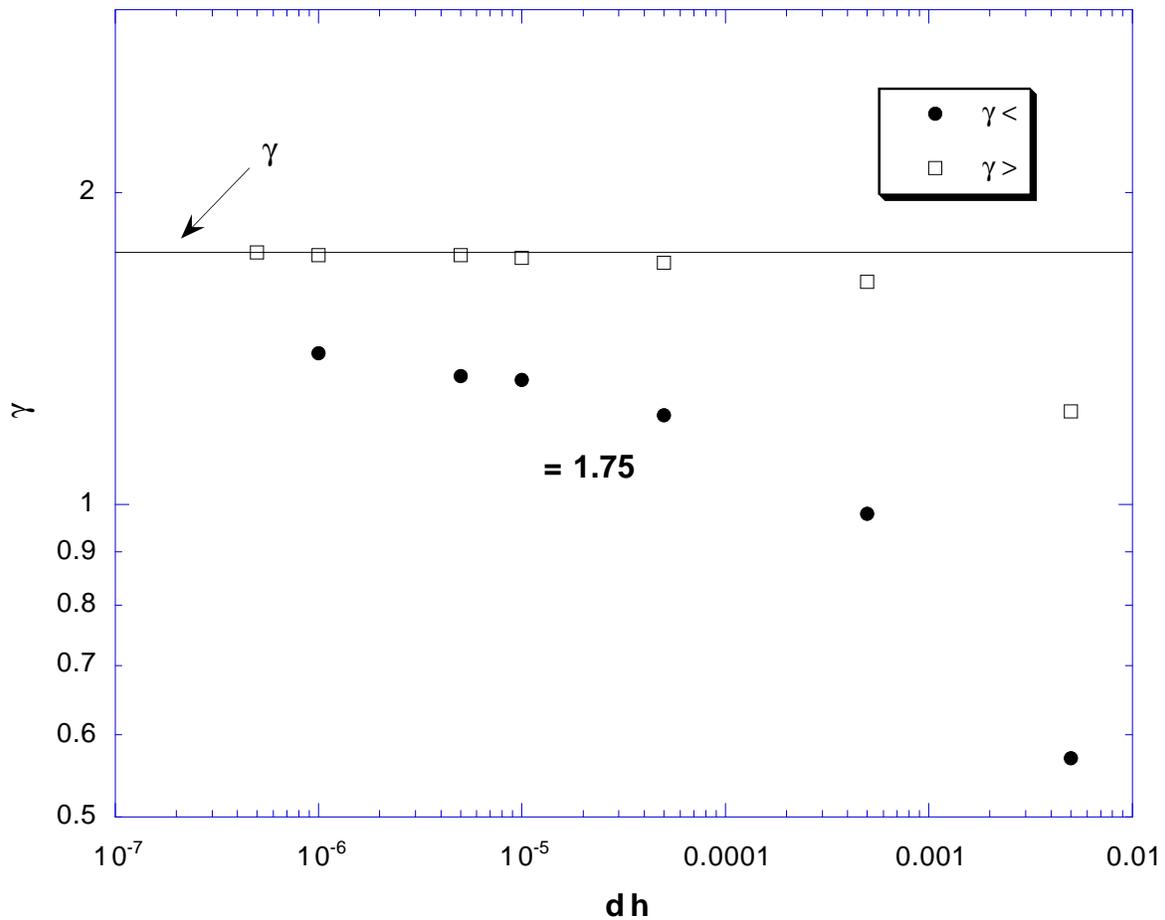,height=4.8in}}
   \caption{The susceptibility exponent $\gamma$ above and below the 
   critical temperature as a function of the 
   grid spacing $dh$. The horizontal line denotes the $2D$ Ising 
   exponent of $\gamma =1.75$.}
    \label{fig:gammaLimit}
\end{figure}

The calculation of $\chi_{T}$ requires special consideration as it is 
dependent on the size of the ``grid'' $\Delta h$ used for the 
numerical differentiation. Best results for the critical exponent 
above and below the critical temperature are: $\gamma = 1.74 \pm 0.01$ 
and $\gamma\prime = 1.33 \pm 0.03$. The exponents do appear to 
converge to the expected value of $7/4$ as the grid size $dh$ is 
reduced (see Figure \ref{fig:gammaLimit}).

Finally we comment on the validity of the assumption that the retention of 
a small number of states in the eigenfunction expansion of the 
transfer matrix (\ref{transfer2}) suffices to ensure an accurate 
calculation of critical point properties.  The previous results were 
obtained by using only two states in the expansion (\ref{transfer2}), 
i.e.  $n_{s}=2$.  To empirically investigate the inclusion of more 
states, the calculation of the specific heat was repeated with 
$n_{s}=4$.  The results are effectively the same within error 
bounds -- $\alpha = \alpha\prime = 0$ and the amplitude ratio is 
$0.95$.

\begin{figure}
    \centering
    \mbox{\epsfig{file=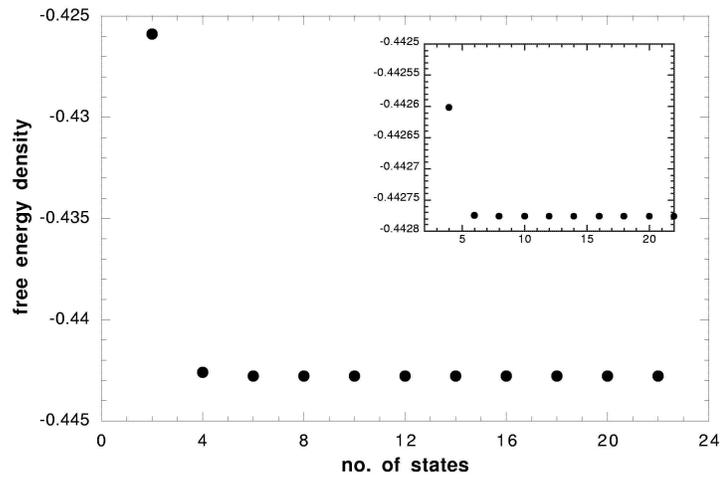,height=4.3in}}
    \caption{The free energy of a chain with a single enhanced bond 
    as a function of $n_{s}$ (i.e.\ the number of states 
    kept in the expansion of the Ginzburg-Landau transfer matrix). The 
    inset shows the detail of $n_{s}=4$ and higher.}
    \label{fig:truncFX}
\end{figure}

There is another perspective from which to explore the effect of 
truncating the Ginzburg-Landau basis.  The {\em lowest order} effect 
of including higher states can be realized by allowing a {\em single} 
bond in the lattice to be represented by the full sum of states in the 
expansion.  We will refer to this as an ``enhanced'' bond.  All other 
bonds are expanded on only the first two ($n_{s}=2$) eigenstates.  We 
can then observe the effect of this enhancement on the free energy 
density as we increase $n_{s}$.  Examining Figure \ref{fig:truncFX} it 
is clear that the effect of enhancement is negligible above $n_{s}=4$.

\end{document}